\documentclass[aps,prb,showpacs,amsmath,amssymb]{revtex4}
\usepackage{graphicx}
\usepackage{bm}

\def\br{ \bm{r} }
\def\bk{ \bm{k} }

\def\bq{ \bm{q} }

\def\bgam{ \bm{\gamma} }
\def\im{ \mathrm{Im}\, }

\begin{document}
\title{Helical states and solitons in noncentrosymmetric superconductors}

\author{K. V. Samokhin}

\affiliation{Department of Physics, Brock University, St. Catharines, Ontario L2S 3A1, Canada}
\date{\today}

\begin{abstract}
We show how the two-band nature of superconductivity in noncentrosymmetric compounds leads to a variety of novel nonuniform superconducting states induced by a magnetic field.
At low fields, a two-band helical state is realized, with a distinctly non-BCS quasiparticle spectrum. At high fields, the superconducting state becomes unstable towards the formation of a lattice of topological phase solitons.
\end{abstract}


\maketitle

\section{Introduction}
\label{sec: Intro}

The recent revival of interest in the properties of two-band (or, more generally, multiband) superconductors has been largely stimulated by the discovery of superconductivity in MgB$_2$ (Ref. \onlinecite{MgB2}). 
Other candidates for multiband superconductivity include nickel borocarbides (Ref. \onlinecite{CGB98}), NbSe$_2$ (Ref. \onlinecite{NbSe2}), CeCoIn$_5$ (Ref. \onlinecite{CeCoIn5}), and also the iron-based 
high-temperature superconductors, see Ref. \onlinecite{Fe-based} for a review. These discoveries have shown that multiband superconductivity, which is characterized by 
a significant difference in the order parameter magnitudes and/or phases in different bands, might be a much more common phenomenon than it was previously thought. 

One important class of multiband superconductors is noncentrosymmetric compounds with a strong spin-orbit (SO) coupling of conduction electrons with the lattice. Since the discovery of superconductivity in CePt$_3$Si 
(Ref. \onlinecite{CePtSi}), the list of noncentrosymmetric superconductors has grown to include dozens of materials, such as UIr (Ref. \onlinecite{Akazawa04}), CeRhSi$_3$ (Ref. \onlinecite{Kimura05}),
CeIrSi$_3$ (Ref. \onlinecite{Sugitani06}), Y$_2$C$_3$ (Ref. \onlinecite{Amano04}), Li$_2$(Pd$_{1-x}$,Pt$_x$)$_3$B (Ref. \onlinecite{LiPt-PdB}), and many others.   
In noncentrosymmetric crystals, the Bloch bands are split by the SO coupling and the Cooper pairing of electrons from different bands is suppressed, resulting in an effectively two-band picture of superconductivity.
Both the spin structure of the bands and the momentum-space topology of the Bloch wavefunctions are nontrivial, which brings about a number of novel properties, 
such as the magnetoelectric effect in the superconducting state,\cite{Lev85,Edel89,Yip02,Fuji05} topologically protected gapless boundary modes and quantum spin Hall effect,\cite{SF09,TYBN09} and the anomalous de Haas-van Alphen and Hall 
effects in the normal state.\cite{Sam09-AP} A comprehensive review of the recent developments in the field can be found in Ref. \onlinecite{Springer-book}. 

One of the most striking features of noncentrosymmetric superconductors is the existence of various unusual nonuniform superconducting states in the presence of a magnetic field $\bm{H}$ (Refs. \onlinecite{MS94,BG02,Agter03,DF03}), 
or even without any field (Ref. \onlinecite{MS08,Sam13}). The best studied example is the helical states, with the order parameter proportional to $e^{i\bq\br}$, where $\bq$ is linear in $\bm{H}$, which originate from 
the deformation of the Bloch bands by the field, see Ref. \onlinecite{Agter-chapter} for a review. The helical states, as well as their nonlinear modifications, such as ``multiple-$\bq$'' states, 
have been previously studied only in some limiting cases, either using a Bardeen-Cooper-Schrieffer (BCS) model with a spin-singlet attraction\cite{DF03,KAS05,AK07} 
or assuming that superconductivity appears only in one of the bands.\cite{Agter03,Sam04} In both cases, the order parameter has one component. 
The purpose of this paper is to develop a phenomenological theory of nonuniform states in noncentrosymmetric superconductors in the general case, fully taking into account the two-component nature of superconductivity in these materials.

Generalization of the BCS theory to the case of two spin-degenerate bands was originally introduced in Ref. \onlinecite{two-band-BCS}. Subsequent work has shown that many properties 
of multiband superconductors differ qualitatively from the single-band case, with the most spectacular features associated with the additional degrees of freedom -- the relative phases of the pair condensates in different bands. 
For example, if the condensate phases have different windings around a vortex core, than the vortex will carry a fractional magnetic flux.\cite{Bab02} In addition to exotic vortices, 
there is another type of topological defects specific to two-band superconductivity, namely phase solitons, in which the relative phase exhibits a kink-like variation by $2\pi$ between its asymptotic mean-field values.\cite{Tanaka01} 
Increasing the number of bands opens up even more intriguing possibilities. For instance, superconducting states that break time-reversal symmetry can exist in ``frustrated'' three-band systems,\cite{SPIS} 
with domain walls separating degenerate ground states.\cite{GB13}
It has been proposed that the phase solitons in two-band superconductors can be dynamically created in nonequilibrium current-carrying states,\cite{GV03} or by the proximity effect with a conventional $s$-wave superconductor,\cite{Vakar12} but 
experimental observation of these effects has remained a challenge. In this paper, we show how the elusive phase solitons can be spontaneously formed in noncentrosymmetric superconductors in a sufficiently strong magnetic field. 

We focus on two-dimensional (2D) noncentrosymmetric superconductors in a parallel field. These systems can be realized experimentally at an interface between two different non-superconducting materials,\cite{interface-SC}
near a doped surface of an insulating crystal,\cite{surface-SC} or near the surface of a topological insulator.\cite{San10} In all these systems the mirror symmetry between two half-spaces separated by an interface is explicitly broken. 
We use the two-band Ginzburg-Landau (GL) formalism\cite{two-band GL} modified
to include first-order gradient terms (the Lifshitz invariants) to catch the effects specific to noncentrosymmetric superconductors.
The paper is organized as follows: In Sec. \ref{sec: two-band picture} we develop a two-band description of noncentrosymmetric superconductors. In Sec. \ref{sec: helical state}, the helical instability in a weak magnetic field is considered
and the quasiparticle spectrum in the helical state is calculated. In Sec. \ref{sec: phase solitons} we discuss the phase solitons and soliton lattices in a strong field. 
Throughout the paper we use the units in which $\hbar=k_B=1$.

\section{Two-band description}
\label{sec: two-band picture}

In order to see how nondegenerate bands are formed in a noncentrosymmetric crystal with the SO coupling, we start with
the following Hamiltonian of noninteracting electrons:
\begin{equation}
\label{H-free}
    H_0=\sum\limits_{\bk,ss'}[\epsilon_0(\bk)\delta_{ss'}+\bgam(\bk)\bm{\sigma}_{ss'}]a^\dagger_{\bk s}a_{\bk s'}.
\end{equation}
Here $s,s'=\uparrow,\downarrow$ are spin indices, $\hat{\bm{\sigma}}$ are the Pauli matrices, $\epsilon_0(\bk)$ is the band dispersion without the SO coupling, and the sum over $\bk$ is restricted to
the first Brillouin zone. The SO coupling of electrons with the crystal lattice is described by $\bgam(\bk)$, which satisfies 
$\bgam(-\bk)=-\bgam(\bk)$. The momentum dependence of the SO coupling is dictated by the requirement that it must be invariant under the crystal symmetry operations, in the following sense: 
if $g$ is any operation from the point group $\mathbb{G}$ of the crystal, then $(g\bgam)(g^{-1}\bk)=\bgam(\bk)$. 
The complete list of representative expressions for $\bgam(\bk)$ for all noncentrosymmetric point groups can be found in Ref. \onlinecite{Sam09-AP}. For instance, in the least symmetric case of a triclinic crystal, i.e. if $\mathbb{G}=\mathbf{C}_1$, 
the direction of $\bgam$ is not related to the crystallographic axes and we have $\gamma_i(\bk)=a_{ij}k_j$, where 
the coefficients $a_{ij}$ form a real $3\times 3$ matrix. In a cubic crystal with $\mathbb{G}=\mathbf{O}$, which describes the point symmetry of
Li$_2$(Pd$_{1-x}$,Pt$_x$)$_3$B, the simplest form compatible with all symmetry requirements is $\bgam(\bk)=\gamma_0\bk$, where $\gamma_0$ is a constant. 
Expressions become more complicated if the point group contains improper elements. For example, for the tetragonal group $\mathbb{G}=\mathbf{C}_{4v}$, which is relevant for CePt$_3$Si, the SO coupling is given by
$\bgam(\bk)=\gamma_0(k_y\hat{\bm{x}}-k_x\hat{\bm{y}})+\gamma_1 k_xk_yk_z(k_x^2-k_y^2)\hat{\bm{z}}$.
Since we focus on the purely 2D case, we can set $\gamma_1=0$. Then the SO coupling takes the Rashba form:
\begin{equation}
\label{Rashba-SOC}
  \bgam(\bk)=\gamma_0(k_y,-k_x),
\end{equation}
which was originally introduced to describe the effects of the absence of mirror symmetry in semiconductor quantum wells.\cite{Rashba60} 

Diagonalizing Eq. (\ref{H-free}) we arrive at the following band dispersion functions:
\begin{equation}
\label{xi-zero-H}
  \xi_\lambda(\bk)=\epsilon_0(\bk)+\lambda|\bgam(\bk)|,
\end{equation}
where $\lambda=\pm$ is called helicity. Therefore, the bands are nondegenerate (except the lines or points where $\bgam=0$), invariant with respect to all operations from $\mathbb{G}$, and also even in $\bk$. 
The last property is a consequence of time reversal symmetry. Indeed, the Bloch states $|\bk,\lambda\rangle$ and $K|\bk,\lambda\rangle$ belong to $\bk$ and $-\bk$, respectively,
and have the same energy. Here $K=i\hat\sigma_2K_0$ is the time reversal operation for spin-$1/2$ particles and $K_0$ is the complex conjugation. Since the bands are nondegenerate, one can write 
$K|\bk,\lambda\rangle=t_\lambda(\bk)|-\bk,\lambda\rangle$, where $t_\lambda(\bk)=-t_\lambda(-\bk)$ is a nontrivial phase factor.\cite{GR01,SC04} The latter is given by
$$
  t_\lambda(\bk)=\lambda\frac{\gamma_x(\bk)-i\gamma_y(\bk)}{\sqrt{\gamma_x^2(\bk)+\gamma_y^2(\bk)}}
$$
for the model described by Eq. (\ref{H-free}).

Let us now introduce an external magnetic field $\bm{H}$. We consider only a 2D superconductor, with the field parallel to its plane, therefore the vector potential and the orbital effects of the field can be neglected. 
Adding the Zeeman interaction $-\mu_B\bm{H}\bm{\sigma}$ to Eq. (\ref{H-free}), where $\mu_B$ is the Bohr magneton, and diagonalizing the resulting Hamiltonian we obtain the following energy eigenvalues: 
$\epsilon_0(\bk)+\lambda|\bgam(\bk)-\mu_B\bm{H}|$. Assuming that the Zeeman energy is small compared with the SO coupling and expanding these eigenvalues to the first order in $\bm{H}$, we have\cite{comment}
\begin{equation}
\label{H_0}
    H_0=\sum_{\bk}\sum_{\lambda=\pm}[\xi_\lambda(\bk)-\lambda\mu_B\hat{\bgam}(\bk)\bm{H}]c^\dagger_{\bk\lambda}c_{\bk\lambda}.
\end{equation}
Therefore, the bands are asymmetrically deformed by the field and no longer even in $\bk$. This allows for the Cooper pairing with a nonzero centre-of-mass momentum to occur, 
leading to a variety of field-induced nonuniform superconducting states, which are discussed in the subsequent sections.   

We assume the following hierarchy of the energy scales: $T_c,\mu_BH\ll\varepsilon_c\ll\max_{\bk}|\bgam(\bk)|\ll\epsilon_F$, where $T_c$ is the superconducting critical temperature, $\varepsilon_c$ is the energy cutoff of the 
pairing interaction, and $\epsilon_F$ is the Fermi energy, which is a good assumption in realistic noncentrosymmetric superconductors.\cite{Springer-book} 
Then the SO-split bands are sufficiently separated for the pairs to form independently in each band, and it is natural to use the basis 
of the helicity band eigenstates to introduce an attractive interaction between electrons in the Cooper channel. Following the standard BCS ideology, we assume that
the pairing interaction is only effective when the quasiparticle momenta are close to the Fermi surfaces and, therefore, the interband pairing, i.e. the formation of the Cooper pairs of electrons with opposite helicities, is suppressed. 
This leads us to the pairing Hamiltonian
\begin{equation}
\label{H int reduced}
    H_{int}=\frac{1}{2{\cal V}}\sum\limits_{\bk\bk'\bq}\sum_{\lambda\lambda'}V_{\lambda\lambda'}(\bk,\bk')c^\dagger_{\bk+\bq,\lambda}c^\dagger_{-\bk,\lambda}c_{-\bk',\lambda'}c_{\bk'+\bq,\lambda'},
\end{equation}
where ${\cal V}$ is the system volume and $V_{\lambda\lambda'}$ is the pairing interaction function. The latter has the form
$$
  V_{\lambda\lambda'}(\bk,\bk')=t_\lambda(\bk)t^*_{\lambda'}(\bk')\tilde V_{\lambda\lambda'}(\bk,\bk'),
$$ 
where $\tilde V_{\lambda\lambda'}(\bk,\bk')$ can be represented as a bilinear combination of the basis functions of irreducible representations of $\mathbb{G}$ and, therefore, is amenable to the usual symmetry analysis.\cite{MinSig-chapter}
Note that the dependence of the pairing interaction on the center-of-mass momentum of the pairs is neglected, which is legitimate since $|\bq|$ is small compared to the Fermi momenta in the two bands.
The simplest model corresponds to isotropic basis functions of the unit representation, with
\begin{equation}
\label{isotropic-pairing}
  \tilde V_{\lambda\lambda'}(\bk,\bk')=-V_{\lambda\lambda'}.
\end{equation}
The strength of intraband pairing is described by the coupling constants $V_{++}$ and $V_{--}$, while that of interband pairing by $V_{+-}=V_{-+}$.

In a uniform superconducting state, using the mean-field approximation to decouple the interaction in Eq. (\ref{H int reduced}) we arrive at the following expression:
\begin{equation}
\label{H_MF}
  H_{int}=\frac{1}{2}\sum_{\bk,\lambda}\left[\Delta_\lambda(\bk)c_{\bk\lambda}^\dagger c_{-\bk\lambda}^\dagger+\Delta_\lambda^*(\bk)c_{-\bk\lambda}c_{\bk\lambda}\right],
\end{equation}
where $\Delta_\lambda$ is the gap function in the $\lambda$th band (we have omitted a $c$-number term on the right-hand side of the last equation). In the isotropic pairing model, see Eq. (\ref{isotropic-pairing}), the gap functions
have the form
\begin{equation}
  \Delta_\lambda(\bk)=t_\lambda(\bk)\eta_\lambda,
\end{equation}
where $\eta_\lambda=|\eta_\lambda|e^{i\varphi_\lambda}$ play the role of the superconducting order parameters.

The model defined by Eqs. (\ref{H_0}) and (\ref{H int reduced}) is formally similar to the two-band BCS theory,\cite{two-band-BCS} 
whose applications to MgB$_2$, iron-based superconductors, and other systems have been extensively studied recently. The order parameter is represented by two
complex functions $\eta_+$ and $\eta_-$. An obvious difference is that in our case the 
two bands are nondegenerate and the gap functions contain the phase factors $t_\lambda(\bk)$. However, these phase factors do not affect the bulk observable quantities which are determined by the quasiparticle excitations, 
such as spin susceptibility and electronic specific heat, nor do they enter the GL free energy, which is expressed in terms of $\eta_+$ and $\eta_-$. 

The order parameters in the helicity band representation are related to the singlet ($\psi$) and ``protected'' triplet ($d$) components in the spin representation as follows:
$\psi=-(\eta_++\eta_-)/2$ and $d=-(\eta_+-\eta_-)/2$ (Ref. \onlinecite{Springer-book}). In particular, in the case of a BCS-like point attraction all the coupling constants in Eq. (\ref{isotropic-pairing}) are the same: 
$V_{++}=V_{--}=V_{+-}$, and we have $\eta_+=\eta_-$, i.e. the pairing is isotropic singlet. In the opposite case of a ``single-band'' model, either $\eta_+$ or $\eta_-$ is zero. In general, however, both components are nonzero
and different, and can also depend on coordinates. In the spin representation this translates into an order parameter having both singlet and triplet components. Nonuniform superconducting states are most efficiently treated using 
the GL formalism.

\subsection{Ginzburg-Landau functional}
\label{sec: GL functional}

The phenomenological GL functional of a noncentrosymmetric superconductor can be obtained in the standard fashion by expanding the free energy (or, more precisely, the difference between the free energies in the superconducting and normal states) 
in powers of $\eta_+$ and $\eta_-$ and their gradients and keeping all terms allowed by symmetry. 
In the case of a 2D noncentrosymmetric superconductor in a parallel magnetic field $\bm{H}$ we have the following expression for the GL free energy density:
\begin{equation}
\label{F_GL}
  F=\sum_{\lambda=\pm} F_\lambda+F_m.
\end{equation}
The intraband contributions are given by
\begin{equation}
\label{F-pm}
  F_\lambda=\alpha_\lambda|\eta_\lambda|^2+\beta_\lambda|\eta_\lambda|^4+K_\lambda|\bm{\nabla}\eta_\lambda|^2+\tilde K_\lambda\im\left[\eta_\lambda^*(\bm{H}\times\bm{\nabla})_z\eta_\lambda\right]+L_\lambda H^2|\eta_\lambda|^2,
\end{equation}
while
\begin{equation}
\label{F-mix}
  F_m=\gamma_m(\eta_+^*\eta_-+\eta_-^*\eta_+)
\end{equation}
is the band-mixing term describing the ``Josephson coupling'' of the two order parameters, which originates in the interband pairing terms in Eq. (\ref{H int reduced}). The free energy (\ref{F_GL}) yields the following expression for the supercurrent:
\begin{equation}
\label{supercurrent}
  \bm{j}=-4e\sum_\lambda K_\lambda\im(\eta_\lambda^*\bm{\nabla}\eta_\lambda)+2e\sum_\lambda\tilde K_\lambda (\bm{H}\times\hat{\bm{z}})|\eta_\lambda|^2,
\end{equation}
where $e$ is the absolute value of the electron charge.

The first three terms in Eq. (\ref{F-pm}) are the usual GL uniform and gradient terms.  The temperature dependence enters through the coefficients $\alpha_\lambda$, which have the form $\alpha_\lambda=a_\lambda(T-T_\lambda)$,
where $a_\lambda>0$ and $T_\lambda$ is the transition temperature the $\lambda$th band would have at zero field in the absence of any interband coupling. The fourth term, sometimes called the Lifshitz invariant, 
is linear both in gradients and in the magnetic field, and is specific to noncentrosymmetric superconductors. Its origin can be traced to the deformation of the helicity bands by the field, see Eq. (\ref{H_0}). Microscopic 
derivation of the Lifshitz invariant can be found in Ref. \onlinecite{Sam04}.
It is the Lifshitz invariants that are responsible for unusual nonuniform states created by the field, see Secs. \ref{sec: helical state} and \ref{sec: phase solitons} below. 
The last term in Eq. (\ref{F-pm}) describes the paramagnetic suppression of the critical temperature, which is due to the change in the paramagnetic susceptibility 
in the superconducting state compared with the normal state. 

Microscopic theory yields expressions for the coefficients in the GL functional in terms of the Fermi-surface averages of the order parameter, the Fermi velocity, 
and the SO coupling direction $\hat{\bgam}$, see Ref. \onlinecite{MinSig-chapter}. Assuming isotropic bands, one has the following order-of-magnitude estimates:
\begin{equation}
\label{coeff-estimates}
  K_\lambda\sim\frac{N_{F,\lambda}}{T_{c0}^2}v_{F,\lambda}^2,\quad \tilde K_\lambda\sim\frac{N_{F,\lambda}}{T_{c0}^2}\mu_Bv_{F,\lambda},\quad L_\lambda\sim\frac{N_{F,\lambda}}{T_{c0}^2}\mu_B^2,
\end{equation}
where $N_{F,\lambda}$ is the Fermi-level DoS and $v_{F,\lambda}$ is the Fermi velocity in the $\lambda$th band, while $T_{c0}$ is the superconducting critical temperature at zero field.

One can recover the single-band and the singlet BCS limits from the general GL functional (\ref{F_GL}) as follows. The single-band case corresponds to the absence of the interband coupling, i.e. $\gamma_m=0$. Assuming $T_+>T_-$, 
superconductivity appears only in the ``$+$'' band, i.e. $\eta_-=0$, at least near the critical temperature. In the BCS case, $\eta_+=\eta_-=\eta$, and one obtains a one-component GL free energy in terms of $\eta$, with the Lifshits 
invariant.

\section{Low fields: Helical state}
\label{sec: helical state}

The critical temperature, $T_c(H)$, of the second-order superconducting phase transition is obtained by solving the linearized GL equations. Assuming $\bm{H}=H\hat{\bm{y}}$ and anticipating a possible helical 
instability modulated perpendicularly to the field, we seek the order parameters in the form $\eta_\lambda(\br)=\eta_\lambda e^{iqx}$. Then from the free energy (\ref{F_GL}) it follows that
\begin{equation}
\label{linear-GL}
  \left(\begin{array}{cc}
        a_+(T-\tilde T_+) & \gamma_m \\
        \gamma_m & a_-(T-\tilde T_-)
  \end{array}\right)
  \left(\begin{array}{c}
         \eta_+ \\ \eta_-
        \end{array}\right)=0,
\end{equation}
where $\tilde T_\lambda=T_\lambda-(K_\lambda q^2-\tilde K_\lambda Hq+L_\lambda H^2)/a_\lambda$. Setting the determinant of the matrix on the left-hand side to zero, one arrives at the following expression for the critical temperature 
as a function of the magnetic field and the helical modulation wavevector:
\begin{equation}
\label{T_c-Hq}
    T_c(H,q)=\frac{\tilde T_++\tilde T_-}{2}+\sqrt{\left(\frac{\tilde T_+-\tilde T_-}{2}\right)^2+\frac{\gamma_m^2}{a_+a_-}}.
\end{equation}
The actual transition temperature $T_c(H)$ and the modulation wavevector are obtained by maximizing the above expression with respect to $q$.

At zero field the maximum critical temperature is achieved in the uniform state, i.e. at $q=0$, and is given by
\begin{equation}
\label{T_c0}
  T_{c0}\equiv T_c(H=0)=\frac{T_++T_-}{2}+\sqrt{\left(\frac{T_+-T_-}{2}\right)^2+\frac{\gamma_m^2}{a_+a_-}}.
\end{equation}
At a small but nonzero field, we seek the wavevector in the form $q\propto H$. Expanding Eq. (\ref{T_c-Hq}) in powers of $H$ and maximizing with respect to $q$, we obtain:
\begin{equation}
\label{q-H}
  q=\frac{A_1}{2A_0}H,
\end{equation}
which corresponds to 
\begin{equation}
\label{T_c-H}
  T_c(H)=T_{c0}-\left(A_2-\frac{A_1^2}{4A_0}\right)H^2.
\end{equation}
Here
\begin{eqnarray}
\label{As}
  && A_0=\frac{1+r}{2}\frac{K_+}{a_+}+\frac{1-r}{2}\frac{K_-}{a_-},\nonumber\\
  && A_1=\frac{1+r}{2}\frac{\tilde K_+}{a_+}+\frac{1-r}{2}\frac{\tilde K_-}{a_-},\\
  && A_2=\frac{1+r}{2}\frac{L_+}{a_+}+\frac{1-r}{2}\frac{L_-}{a_-},\nonumber
\end{eqnarray}
and 
$$
  r=\frac{T_+-T_-}{\sqrt{(T_+-T_-)^2+4\gamma_m^2/a_+a_-}}.
$$
Thus we see that, similarly to the single-component helical state,\cite{Agter-chapter} the superconducting order parameter is nonuniform, with the modulation wavevector linearly proportional to the field. 
The suppression of the critical temperature is quadratic in $H$, which is typical of a paramagnetic pair breaking. According to Eq. (\ref{T_c-H}), the pair breaking is weakened in the presence of the helical instability. 
We would like to stress that, in contrast to the Larkin-Ovchinnikov-Fulde-Ferrell nonuniform state,\cite{FFLO} which only exists in paramagnetically-limited superconductors at sufficiently strong magnetic fields,
the helical state appears at an arbitrarily weak field.

The last term on the right-hand side of Eq. (\ref{F-pm}), which describes the paramagnetic suppression of superconductivity, is crucial for maintaining the stability of the system at $H=0$.
If this term were not included, the linear in gradient terms would result in the critical temperature unphysically increasing as a function of the field. To avoid this, one has to assume that $A_2\geq A_1^2/4A_0$. 

Despite the fact that the order parameter in the helical state is proportional to $e^{iqx}$, the supercurrent is equal to zero. Indeed, Eq. (\ref{supercurrent}) yields the following expression:
\begin{equation}
\label{current-x}
  j_x=-4e\sum_\lambda K_\lambda|\eta_\lambda|^2\nabla_x\varphi_\lambda+2eH\sum_\lambda \tilde K_\lambda|\eta_\lambda|^2,
\end{equation}
where $\varphi_\lambda=qx$, and $j_y=0$. Substituting Eq. (\ref{q-H}), we obtain:
\begin{equation}
  j_x=-2eH\sum_\lambda\left(\frac{A_1}{A_0}K_\lambda-\tilde K_\lambda\right)|\eta_\lambda|^2=0.
\end{equation}
Here we used the ratio of the order parameters at zero field, $\eta_-/\eta_+=-\gamma_m/a_-(T-T_-)$, which follows from Eq. (\ref{linear-GL}). That the current must vanish can also be understood using the following simple argument. Uniform supercurrent 
is obtained by differentiating the total free energy with respect to a uniform vector potential: $j_x=-(c/{\cal V})\partial{\cal F}/\partial A_x$, where $c$ is the speed of light. Due to the gauge invariance, we have
$\partial{\cal F}/\partial A_x=-(2e/c)\partial{\cal F}/\partial q=0$ and, therefore, $j_x=0$, because the free energy has a minimum at the equilibrium value of the helical modulation.

\subsection{General case}
\label{sec: helical-general}

Although the above calculations apply in the case of a 2D isotropic noncentrosymmetric superconductor, they can be straightforwardly extended to an arbitrary in-plane symmetry.
In the general case, the Lifshitz invariant is bilinear in both the order parameter gradients and the magnetic field and Eq. (\ref{F-pm}) is replaced by
\begin{equation}
\label{F-pm-general}
  F_\lambda=\alpha_\lambda|\eta_\lambda|^2+\beta_\lambda|\eta_\lambda|^4+K_{\lambda,ij}(\nabla_i\eta_\lambda^*)(\nabla_j\eta_\lambda)+\tilde K_{\lambda,ij}\im(\eta_\lambda^*\nabla_i\eta_\lambda)H_j+L_{\lambda,ij}H_iH_j|\eta_\lambda|^2,
\end{equation}
where $i,j=x,y$, and the Einstein summation convention is assumed. The matrices $\hat K_{\lambda}$ and $\hat L_\lambda$ are symmetric, while $\hat{\tilde K}_{\lambda}$ is neither symmetric nor antisymmetric, in general. 
The isotropic case is recovered when $K_{\lambda,ij}=K_\lambda\delta_{ij}$, $L_{\lambda,ij}=L_\lambda\delta_{ij}$, and $\tilde K_{\lambda,ij}=-e_{zij}\tilde K_\lambda$. 
In the least symmetric case of $\mathbb{G}=\mathbf{C}_1$ all elements of the matrices $\hat K_{\lambda}$, $\hat{\tilde K}_{\lambda}$, and $\hat L_\lambda$ are nonzero. 

We seek the order parameter in the form $\eta_\lambda(\br)=\eta_\lambda e^{i\bq\br}$ and obtain for the critical temperature the same expression as Eq. (\ref{T_c-Hq}).
The only difference is that $\tilde T_\lambda$ are now given by
$$
  \tilde T_\lambda=T_\lambda-\frac{1}{a_\lambda}(K_{\lambda,ij}q_iq_j+\tilde K_{\lambda,ij}q_iH_j+L_{\lambda,ij}H_iH_j).
$$
Maximizing the critical temperature with respect to $\bq$, we obtain:
\begin{equation}
\label{q-H-gen}
  q_i=-\frac{1}{2}(\hat A_0^{-1})_{ij}A_{1,jk}H_k
\end{equation}
and 
\begin{equation}
\label{Tc-H-gen}
  T_c(\bm{H})=T_{c0}-\left[A_{2,ij}-\frac{1}{4}(\hat A_1^T\hat A_0^{-1}\hat A_1)_{ij}\right]H_iH_j,
\end{equation}
where 
\begin{eqnarray*}
  && \hat A_0=\frac{1+r}{2}\frac{\hat K_+}{a_+}+\frac{1-r}{2}\frac{\hat K_-}{a_-},\\
  && \hat A_1=\frac{1+r}{2}\frac{\hat{\tilde K}_+}{a_+}+\frac{1-r}{2}\frac{\hat{\tilde K}_-}{a_-},\\
  && \hat A_2=\frac{1+r}{2}\frac{\hat L_+}{a_+}+\frac{1-r}{2}\frac{\hat L_-}{a_-}.
\end{eqnarray*}
Therefore, the critical temperature is suppressed by the field, quadratically in $\bm{H}$, but the helical modulation wavevector is no longer perpendicular to the field, in general. 
The supercurrent is equal to zero, for the same reason as explained above.

\subsection{Density of quasiparticle states}
\label{sec: DoS}

The helical states can be observed, for instance, in tunneling experiments, which measure the density of states (DoS) of quasiparticle excitations. In this subsection we calculate the quasiparticle spectrum in the state
with $\eta_\lambda(\br)=|\eta_\lambda|e^{i\bq\br}$, by solving the Bogoliubov-de Gennes (BdG) equations independently in each band. 
Derivation of the BdG equations for a noncentrosymmetric superconductor is presented in Appendix \ref{sec: BdG-helical}.

The helical state order parameter in the momentum representation is given by $\eta_\lambda(\bm{p})={\cal V}\delta_{\bm{p},\bq}\eta_\lambda$, and the BdG Hamiltonian, 
see Eq. (\ref{H-BdG-transformed}), takes the following form:  
\begin{equation}
\label{H-BdG-helical}
  \tilde{\cal H}_\lambda(\bk,\bk')=\left(\begin{array}{cc}
                                \delta_{\bk,\bk'}\Xi_\lambda(\bk) &  \delta_{\bk-\bk',\bq}\eta_\lambda \\
                                \delta_{\bk'-\bk,\bq}\eta^*_\lambda & -\delta_{\bk,\bk'}\Xi_\lambda(-\bk')
                                \end{array}\right),
\end{equation}
where $\Xi_\lambda(\bk)=\xi_\lambda(\bk)-\lambda\mu_B\hat{\bgam}(\bk)\bm{H}$. To make the above Hamiltonian diagonal in $\bk$-space, we perform a unitary transformation 
$\tilde{\cal H}_\lambda\to H_\lambda=W_{\bq}\tilde{\cal H}_\lambda W^\dagger_{\bq}$,
where $W_{\bq}=\exp(-i\bq\hat\br\hat\sigma_3/2)$ and $\hat\br=i\nabla_{\bk}$ is the position operator in $\bk$-space. One can see that the exponentials in $W_{\bq}$ act as finite displacement operators in momentum space, because
$\langle\bk|\exp(i\bm{Q}\hat\br)|\bk'\rangle=\langle\bk|\bk'+\bm{Q}\rangle=\delta_{\bk,\bk'+\bm{Q}}$. After the unitary transformation, the BdG Hamiltonian becomes
$$
  H_\lambda(\bk,\bk')=\delta_{\bk,\bk'}\left(\begin{array}{cc}
                                \Xi_\lambda\left(\bk+\frac{\bq}{2}\right) &  \eta_\lambda \\
                                \eta^*_\lambda & -\Xi_\lambda\left(-\bk+\frac{\bq}{2}\right)
                                \end{array}\right).
$$
In a weak field, the helical modulation wavevector is small and one can expand the last expression in powers of $\bq$ and $\bm{H}$, with the following result:
\begin{equation}
\label{H-BdG-weak-field}
  H_\lambda(\bk,\bk')=\delta_{\bk,\bk'}\left(\begin{array}{cc}
                                \xi_\lambda(\bk)+\Omega_\lambda(\bk) &  \eta_\lambda \\
                                \eta^*_\lambda & -\xi_\lambda(\bk)+\Omega_\lambda(\bk)
                                \end{array}\right),
\end{equation}
where 
$$
  \Omega_\lambda(\bk)=\frac{1}{2}\bm{v}_\lambda(\bk)\bq-\lambda\mu_B\hat{\bgam}(\bk)\bm{H}
$$
describes the field-induced deformation of the Fermi surface and $\bm{v}_\lambda=\nabla_{\bk}\xi_\lambda$ is the band velocity of quasiparticles.

Now we are in a position to calculate the quasiparticle DoS in the the $\lambda$th band:
\begin{equation}
\label{DoS}
  N_\lambda(E)=\frac{1}{2}\frac{1}{\cal V}\sum_{\bk}\sum_{p=\pm}\delta[E-E^{(p)}_\lambda(\bk)],
\end{equation}
where 
$E^{(\pm)}_\lambda(\bk)=\pm\sqrt{\xi_\lambda^2(\bk)+|\eta_\lambda|^2}+\Omega_\lambda(\bk)$ are the eigenvalues of the Hamiltonian (\ref{H-BdG-weak-field}). Near the Fermi surface, one can integrate Eq. (\ref{DoS}) over $\xi_\lambda$ and obtain:
\begin{equation}
\label{DoS-final}
  N_\lambda(E)=N_{F,\lambda}\left\langle\frac{|E-\Omega_\lambda(\bk)|}{\sqrt{[E-\Omega_\lambda(\bk)]^2-|\eta_\lambda|^2}}\right\rangle_{FS},
\end{equation}
where the angular brackets denote the average over the $\lambda$th Fermi surface, restricted by the condition $|E-\Omega_\lambda(\bk)|\geq|\eta_\lambda|$. Since the DoS satisfies the BdG electron-hole symmetry, 
$N_\lambda(E)=N_\lambda(-E)$, we focus only on the upper half of the spectrum, i.e. on $E\geq 0$.

\begin{figure}
    \includegraphics[width=6cm]{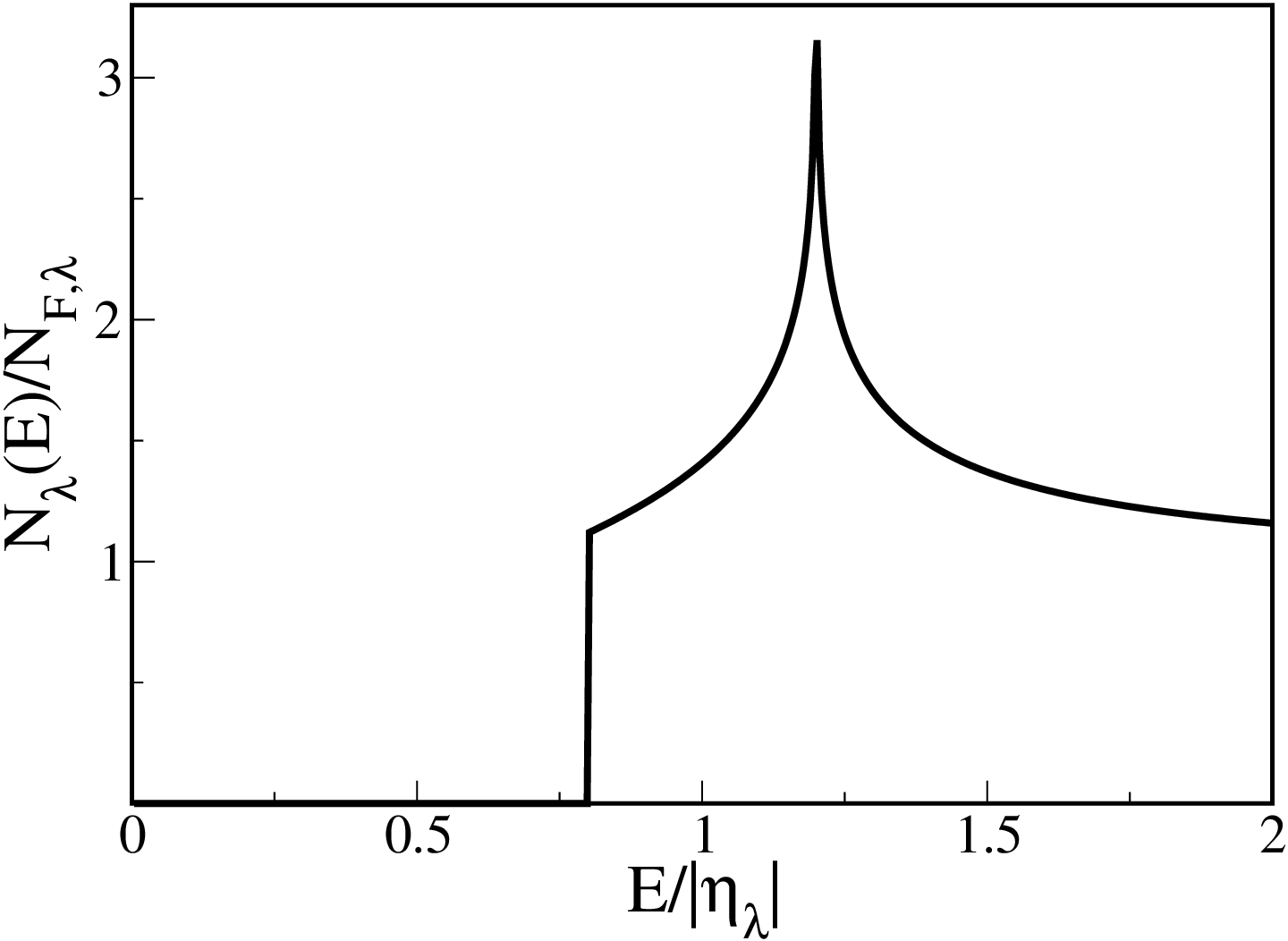}
    \caption{Quasiparticle DoS in the $\lambda$th band, for $\mu_\lambda H/|\eta_\lambda|=0.2$. The gap edge is shifted to $E/|\eta_\lambda|=1-\mu_\lambda H/|\eta_\lambda|$, while 
    the logarithmic singularity occurs at $E/|\eta_\lambda|=1+\mu_\lambda H/|\eta_\lambda|$.}
    \label{fig: DoS}
\end{figure}

\begin{figure}
    \includegraphics[width=6cm]{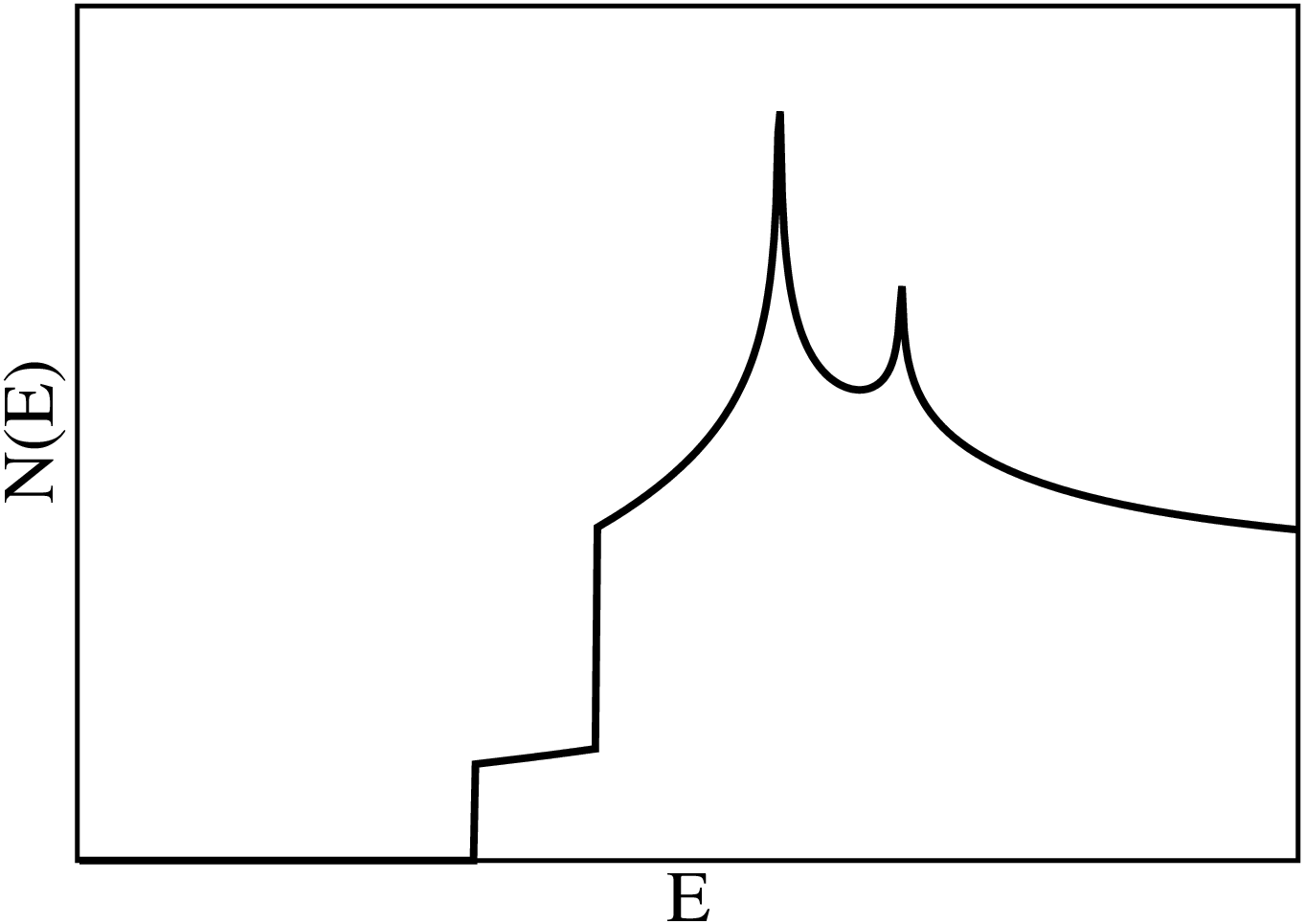}
    \caption{Total quasiparticle DoS, $N(E)=N_+(E)+N_-(E)$, in the two-band helical state.}
    \label{fig: DoS-total}
\end{figure}

To make analytical progress we assume a cylindrical Fermi surface with the Rashba SO coupling, see Eq. (\ref{Rashba-SOC}), and $\bm{H}=H\hat{\bm{y}}$, with $H>0$. Then,
$$
  \Omega_\lambda(\bk)=\mu_\lambda H\hat k_x,\quad \mu_\lambda=\lambda\mu_B+\frac{v_{F,\lambda}A_1}{4A_0}.
$$
It follows from Eqs. (\ref{As}) and (\ref{coeff-estimates}) that both the Zeeman and ``helical'' contributions to $\Omega_\lambda$, described by the first and second terms in $\mu_\lambda$, respectively, are of the same order. 
From Eq. (\ref{DoS-final}) we obtain the following expression for the DoS:
\begin{equation}
\label{DoS-Rashba}
  N_\lambda(E)=N_{F,\lambda}I\left(\frac{E}{|\eta_\lambda|},\frac{\mu_\lambda H}{|\eta_\lambda|}\right),
\end{equation}
where
\begin{equation}
\label{Ixy}
  I(x,y)=\int_0^{2\pi}\frac{d\phi}{2\pi}\frac{|x-y\cos\phi|}{\sqrt{(x-y\cos\phi)^2-1}}.
\end{equation}
The angular integration in here is restricted by the condition $|x-y\cos\phi|\geq 1$. We calculated the last integral numerically, with the result presented in Fig. \ref{fig: DoS}. 
This plot reveals two prominent features. First, the gap in the quasiparticle spectrum is given by 
$|\eta_\lambda|-\mu_\lambda H$, i.e. it is smaller in the helical state than in the uniform state. Second, the inverse-square-root DoS singularity, which is a hallmark of the BCS theory, 
is replaced by a much weaker singularity at $E\to|\eta_\lambda|+\mu_\lambda H$. A straightforward analytical evaluation of Eq. (\ref{Ixy}) gives
\begin{equation}
\label{DoS-edge}
  \frac{N_\lambda(E)}{N_{F,\lambda}}=\frac{1}{2}\sqrt{\frac{|\eta_\lambda|}{\mu_\lambda H}},
\end{equation}
immediately above the gap edge, and
\begin{equation}
\label{DoS-peak}
  \frac{N_\lambda(E)}{N_{F,\lambda}}=\frac{1}{2\pi}\sqrt{\frac{|\eta_\lambda|}{\mu_\lambda H}}\ln\frac{\mu_\lambda H}{|E-(|\eta_\lambda|+\mu_\lambda H)|},
\end{equation}
near the peak. Tunneling experiments probe the total DoS, which is given by $N(E)=N_+(E)+N_-(E)$ and sketched in Fig. \ref{fig: DoS-total}.

\section{High fields: soliton lattice}
\label{sec: phase solitons}

In this section we show that a soliton-like texture is spontaneously formed in the superconducting state above a certain critical magnetic field.
Phase solitons are the simplest topological defects that can exist in a two-band superconductor. In a phase soliton the relative phase of the two order parameters exhibits a kink-like 
variation of $2\pi$, approaching its mean-field value at infinity.\cite{Tanaka01} According to Sec. \ref{sec: two-band picture}, noncentrosymmetric superconductors can be viewed as two-band systems and, therefore, are expected to support phase
solitons.

We assume the in-plane isotropy and set $\bm{H}=H\hat{\bm{y}}$. To make analytical progress, we work at zero temperature and employ the London approximation, in which the gap magnitudes are constant. 
Considering a planar texture perpendicular to the $x$ axis, the order parameter components are given by $\eta_\lambda(\br)=|\eta_\lambda|e^{i\varphi_\lambda(x)}$. 
The supercurrent in this state is given by Eq. (\ref{current-x}). The current conservation implies that $j_x=\mathrm{const}$, where the value of the constant is set by external sources and can be assumed to be zero. 
The condition $j_x=0$ allows one to express the gradients of $\varphi_+$ and $\varphi_-$ in terms of the gradient of the relative phase $\theta=\varphi_+-\varphi_-$:
\begin{equation}
\label{phi-grads}
  \nabla_x\varphi_+=\frac{1}{1+\rho}\nabla_x\theta+q,\quad 
  \nabla_x\varphi_+=-\frac{\rho}{1+\rho}\nabla_x\theta+q,
\end{equation}
where 
$$
  \rho=\frac{K_+|\eta_+|^2}{K_-|\eta_-|^2},\quad q=\frac{H}{2}\frac{\sum_\lambda \tilde K_\lambda|\eta_\lambda|^2}{\sum_\lambda K_\lambda|\eta_\lambda|^2}.
$$
It follows from Eq. (\ref{phi-grads}) that one can have two very different physical situations, depending on whether the phases $\varphi_+$ and $\varphi_-$ are locked together, with $\theta$ taking a constant value throughout the system,
or they are allowed to vary independently, leading to a spatially-nonuniform $\theta$. While the former case corresponds to the helical states considered in Sec. \ref{sec: helical state}, in the present section we focus on the latter possibility.

Substituting Eq. (\ref{phi-grads}) into Eq. (\ref{F_GL}), we obtain: $F=2Kf+(...)$. Here the ellipsis denote the terms which depend only on $|\eta_+|$ and $|\eta_-|$, while 
the contributions containing the relative phase have the form
\begin{equation}
\label{F-theta}
  f=\frac{1}{2}(\nabla_x\theta)^2-h(\nabla_x\theta)+\frac{\gamma_m|\eta_+||\eta_-|}{K}\cos\theta,
\end{equation}
where
\begin{equation}
\label{h-def}
  h=uH,\qquad u=\frac{1}{2}\left(\frac{\tilde K_+}{K_+}-\frac{\tilde K_-}{K_-}\right),
\end{equation}
and 
$$
  K=\frac{K_+K_-|\eta_+|^2|\eta_-|^2}{K_+|\eta_+|^2+K_-|\eta_-|^2}.
$$
According to Eq. (\ref{coeff-estimates}), we have the following estimate: $u\sim\mu_B(v_{F,+}^{-1}-v_{F,-}^{-1})$. Without loss of generality we assume that $u>0$.

For concreteness we consider only the case $\gamma_m<0$ (interband attraction), which means that if the relative phase is constant then it is equal to zero. The results for $\gamma_m>0$ are obtained by shifting the whole phase 
texture by $\pi$. For the free energy difference, $\delta f$, between the state with a nonuniform $\theta(x)$ and the state 
with $\theta=0$ we obtain:
\begin{equation}
\label{delta f}
  \delta f=\frac{1}{2}(\nabla_x\theta)^2+V_0(1-\cos\theta)-h(\nabla_x\theta),
\end{equation}
where $V_0=|\gamma_m||\eta_+||\eta_-|/K$. Variational minimization of the last expression yields the following nonlinear differential equation for the relative phase:
\begin{equation}
\label{theta-eq}
  \nabla_x^2\theta-V_0\sin\theta=0.
\end{equation}
In addition to the uniform solution $\theta=0$, this equation also has various nonuniform ones corresponding to phase solitons or soliton lattices. For example, a single soliton connects $\theta=0$ at $x\to-\infty$ and
$\theta=2\pi$ at $x\to\infty$ and has the following explicit form: $\theta(x)=\pi+2\arcsin[\tanh(x/\xi_s)]$, where $\xi_s=1/\sqrt{V_0}$ is the soliton width. 

While the last term in Eq. (\ref{delta f}) is a full derivative and does not contribute to the equation of motion, it does affect the free energy. In fact it is easy to see that this term is responsible for a
phase transition in the system, which is controlled by the external field. Qualitatively, the parameter $h$, which is proportional to the magnetic field, provides a bias favoring a nonzero average gradient of the relative phase. 
When this bias becomes large enough to overcome the energy cost of creating the solitons, the latter will be spontaneously formed in the system.
Denoting the single-soliton energy by $\epsilon_1$ and assuming that the density of solitons $n_s$ is low, i.e. the spacing between them is much greater than $\xi_s$, 
one can expand the difference between the total free energies of the system with and without solitons in powers of $n_s$: 
\begin{equation}
\label{delta-f-n_s}
  \lim_{L_x\to\infty}\frac{1}{L_x}\int_0^{L_x} dx\;\delta f=(\epsilon_1-2\pi h)n_s+...,
\end{equation}
where $L_x$ is the length of the system in the $x$ direction. The neglected terms, denoted by the ellipsis, take into account the interaction between the solitons. One can see that at $h>h_s$, where 
\begin{equation}
\label{h_s}
  h_s=\frac{\epsilon_1}{2\pi},
\end{equation}
the leading term in Eq. (\ref{delta-f-n_s}) becomes negative, resulting in the proliferation of phase solitons. 

\begin{figure}
    \includegraphics[width=6cm]{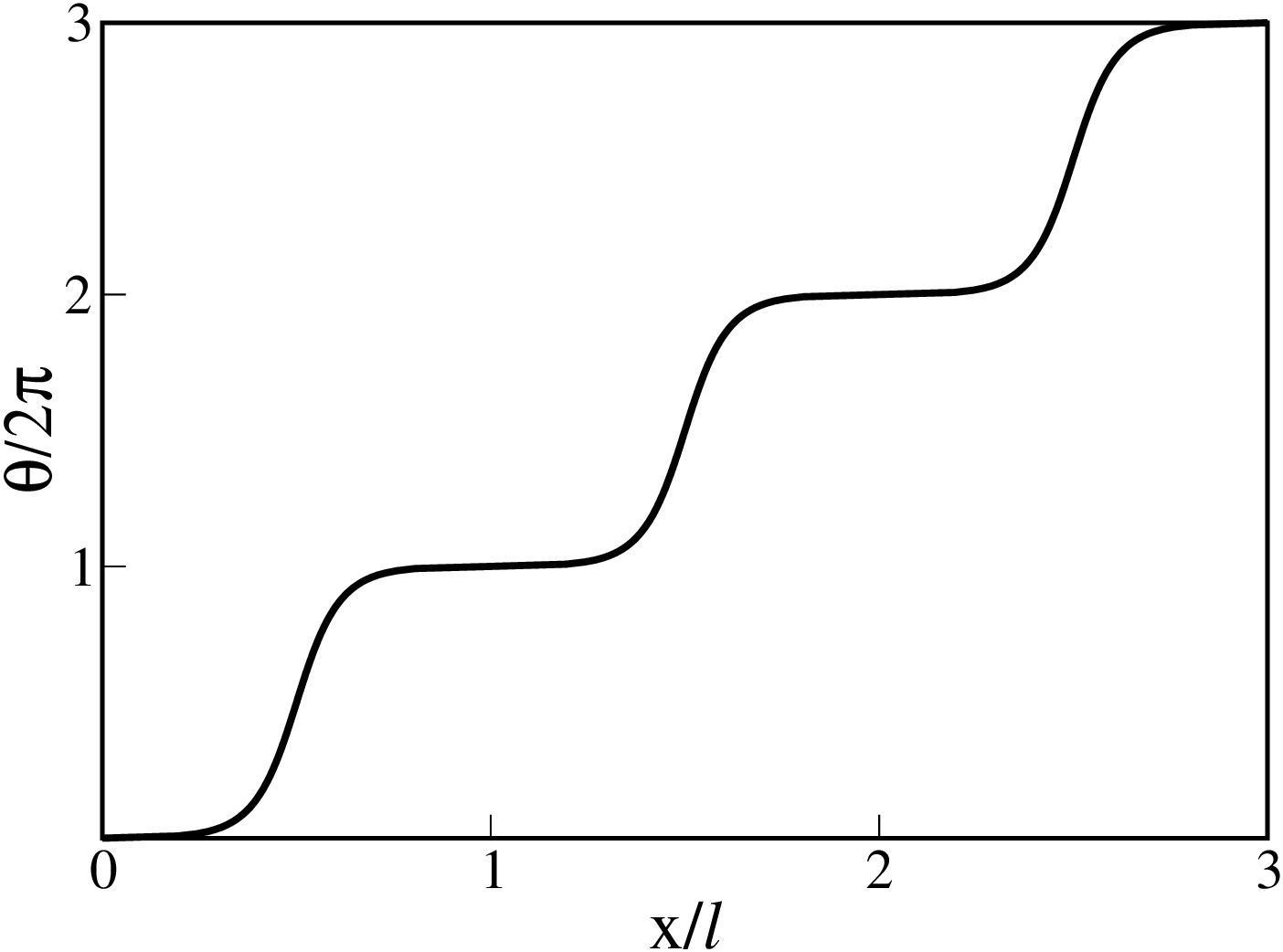}
    \caption{Soliton lattice texture in the phase difference, $\theta(x)$, between the two bands, at high magnetic fields, $H>H_s$ ($\ell$ is the soliton lattice period).}
    \label{fig: theta}
\end{figure}

The transition between the state with a uniform relative phase and the soliton lattice state is mathematically similar to the commensurate-incommensurate transition
of noble gas atoms adsorbed on a periodic substrate.\cite{ChLub-book} Banishing the technical details of the solution to Appendix \ref{sec: SL-transition}, here we present only the results.
The critical magnetic field above which the phase soliton lattice is formed is given by   
\begin{equation}
\label{H_s-final}
  H_s=\frac{4}{\pi}\sqrt{\frac{|\gamma_m||\eta_+||\eta_-|}{Ku^2}},
\end{equation}
see Eqs. (\ref{h_s-V_0}) and (\ref{h-def}). Note that the critical field of the superconducting transition in 2D noncentrosymmetric systems diverges at $T\to 0$, see Refs. \onlinecite{BG02} and \onlinecite{DF03}, therefore, 
the soliton instability of the superconducting state is always realized at strong enough fields. According to Eq. (\ref{l near h_s}), the spacing between solitons grows logarithmically at $H\to H_s+0$:
\begin{equation}
\label{spacing near H_s}
  \frac{\ell(H)}{\xi_s}\simeq 2\ln\frac{H_s}{H-H_s}.
\end{equation}
The relative phase of the order parameter components in the soliton lattice at $H>H_s$ is shown in Fig. \ref{fig: theta}. 

It follows from Eq. (\ref{phi-grads}) that at low fields, $H<H_s$, when the relative phase is uniform, 
the order parameters in the two bands are still nonuniform and given by $|\eta_\lambda|e^{iqx}$. Thus the two-band helical state studied in Sec. \ref{sec: helical state} is recovered. 

\section{Conclusions}
\label{sec: conclusions}

We have developed a general theory of the helical instability in two-dimensional noncentrosymmetric superconductors, taking into account the two-component nature of the order parameter in these systems. 
We have found that the paramagnetic pair breaking is weakened in the presence of the helical modulation. The quasiparticle DoS in the helical state is significantly different from that in a uniform state, showing field-dependent
gap edges and logarithmic singularities, which could be probed in tunneling expreriments.

We have also found a novel type of field-induced nonuniform superconducting state, namely the lattice of phase solitons. Unlike the previously studied nonuniform states in noncentrosymmetric superconductors,
the phase solitons appear only in the two-component model. In contrast to the phase solitons in centrosymmetric two-band superconductors, which are difficult to create, 
the soliton instability predicted in this paper is always present at sufficiently strong magnetic fields. The transition into the soliton state takes place when the bias provided by the Lifshitz invariants in the GL free energy, 
which are unique to noncentrosymmetric systems, overcomes the energy cost
of creating a soliton. This phase transition should show up as a feature in the high-field low-temperature portion of the phase diagram. One can also expect that the spatial inhomogeneity of the order parameters will result
in a qualitative modification of the quasiparticle spectrum, similar to the soliton bound states in centrosymmetric two-band superconductors,\cite{Sam-ABS-2-band} which could be studied by tunneling. This and other issues, such as the fate
of the soliton state at finite temperatures or in the presence of the orbital effects of magnetic field, will be studied elsewhere.

\acknowledgments

This work was supported by a Discovery Grant from the Natural Sciences and Engineering Research Council of Canada.

\appendix

\section{BdG equations in noncentrosymmetric superconductors}
\label{sec: BdG-helical}

The BdG equations, which determine the quasiparticle spectrum in an arbitrary nonuniform superconducting state, can be derived using the standard mean-field approach of the BCS theory, with some modifications 
pertinent to noncentrosymmetric superconductors. The starting point is the Hamiltonian $H=H_0+H_{int}$, where $H_0=\sum_{\bk\lambda}\Xi_\lambda(\bk)c^\dagger_{\bk\lambda}c_{\bk\lambda}$ is the noninteracting part, with 
$\Xi_\lambda(\bk)=\xi_\lambda(\bk)-\lambda\mu_B\hat{\bgam}(\bk)\bm{H}$ being the band dispersion function deformed by the magnetic field, see Eq. (\ref{H_0}), 
and $H_{int}$ is the pairing interaction given by Eq. (\ref{H int reduced}). Decoupling the latter in the mean-field approximation, we obtain:
\begin{eqnarray}
\label{H_int-mf}
  H_{int}=\frac{1}{2{\cal V}}\sum_{\bk\bq\lambda}\left[\Delta_\lambda(\bk,\bq)c_{\bk+\bq,\lambda}^\dagger c_{-\bk\lambda}^\dagger+\Delta_\lambda^*(\bk,\bq)c_{-\bk\lambda}c_{\bk+\bq,\lambda}\right]\nonumber\\
    -\frac{1}{2{\cal V}}\sum_{\bk\bk'\bq}\sum_{\lambda\lambda'}\Delta^*_\lambda(\bk,\bq)V^{-1}_{\lambda\lambda'}(\bk,\bk')\Delta_{\lambda'}(\bk',\bq).
\end{eqnarray}
Here $\Delta_\lambda$ is the gap function in the $\lambda$th band, which satisfies the self-consistency equation
$$
  \Delta_\lambda(\bk,\bq)=\sum_{\bk'\lambda'}V_{\lambda\lambda'}(\bk,\bk')\left\langle c_{-\bk',\lambda'}c_{\bk'+\bq,\lambda'}\right\rangle,
$$
and $V^{-1}$ in the second term in Eq. (\ref{H_int-mf}) should be understood as the inverse matrix both in $\bk$- and $\lambda$-spaces. 

Separating the phase factors $t_\lambda(\bk)$ introduced in Sec. \ref{sec: two-band picture}, we have
$V_{\lambda\lambda'}(\bk,\bk')=t_\lambda(\bk)t^*_{\lambda'}(\bk')\tilde V_{\lambda\lambda'}(\bk,\bk')$ and, therefore, $\Delta_\lambda(\bk,\bq)=t_\lambda(\bk)\tilde\Delta_\lambda(\bk,\bq)$, where the $\bk$-dependence of $\tilde\Delta_\lambda$ is 
determined by the basis functions of an irreducible representation of the point group. Assuming an isotropic pairing, which corresponds to the unit representation, we have $\tilde\Delta_\lambda(\bk,\bq)=\eta_\lambda(\bq)$ and 
\begin{equation}
\label{Delta-isotropic}
  \Delta_\lambda(\bk,\bq)=t_\lambda(\bk)\eta_\lambda(\bq).
\end{equation}
Two complex functions $\eta_+$ and $\eta_-$ comprise the order parameter of our noncentrosymmetric superconductor. 

Introducing the Nambu operators in each band, $C_{\bk\lambda}=(c_{\bk\lambda},c^\dagger_{-\bk,\lambda})^\top$, one can write the fermionic part of the mean-field Hamiltonian in the following form:
\begin{equation}
  H_F=\frac{1}{2}\sum_{\bk\bk'\lambda}C^\dagger_{\bk\lambda}{\cal H}_\lambda(\bk,\bk')C_{\bk'\lambda},
\end{equation}
where 
\begin{equation}
\label{H-BdG}
  {\cal H}_\lambda(\bk,\bk')=\left(\begin{array}{cc}
                                \delta_{\bk,\bk'}\Xi_\lambda(\bk) &  {\cal V}^{-1}\Delta_\lambda(\bk',\bk-\bk') \\
                                {\cal V}^{-1}\Delta^*_\lambda(\bk,\bk'-\bk) & -\delta_{\bk,\bk'}\Xi_\lambda(-\bk')
                                \end{array}\right)
\end{equation}
are the matrix elements of the BdG Hamiltonian in momentum representation. Thus the spectrum of the Bogoliubov quasiparticles can be found independently in each band.

The phase factors in the off-diagonal elements of the Hamiltonian (\ref{H-BdG}) can be removed by a unitary transformation: 
$$
  \tilde{\cal H}_\lambda(\bk,\bk')=U_\lambda(\bk){\cal H}_\lambda(\bk,\bk')U^\dagger_\lambda(\bk'),
$$
where
$$
  U_\lambda(\bk)=\left(\begin{array}{cc}
                       1 & 0 \\
                       0 & t_\lambda(\bk)
                       \end{array}\right).
$$
The matrix elements of the transformed Hamiltonian in momentum space have the following form:
\begin{equation}
\label{H-BdG-transformed}
  \tilde{\cal H}_\lambda(\bk,\bk')=\left(\begin{array}{cc}
                                \delta_{\bk,\bk'}\Xi_\lambda(\bk) &  {\cal V}^{-1}\eta_\lambda(\bk-\bk') \\
                                {\cal V}^{-1}\eta^*_\lambda(\bk'-\bk) & -\delta_{\bk,\bk'}\Xi_\lambda(-\bk')
                                \end{array}\right).
\end{equation}
We see that the phase factors $t_\lambda(\bk)$ do not affect the quasiparticle spectrum in a nonuniform superconducting state (as long as there is no impurities or external fields in the diagonal elements of the BdG Hamiltonian). 
In a uniform state, the order parameter has the form $\eta_\lambda(\bq)={\cal V}\delta_{\bq,0}\eta_\lambda$, and the diagonalization of Eq. (\ref{H-BdG-transformed}) yields $\sqrt{\Xi_\lambda^2(\bk)+|\eta_\lambda|^2}$ for the energy of an
excitation with wavevector $\bk$.

\section{Transition into the soliton lattice}
\label{sec: SL-transition}

It is straightforward to check that Eq. (\ref{theta-eq}) has the following integral of motion:
\begin{equation}
\label{int-of-motion}
  {\cal E}=\frac{1}{2}\left(\nabla_x\theta\right)^2-V_0(1-\cos\theta)=\mathrm{const}.
\end{equation}
This admits a simple mechanical analogy: interpreting $\theta$ as a coordinate and $x$ as a time variable, $\cal{E}$ has the meaning of the total energy of a pendulum of mass $m=1$ in the potential $V(\theta)=-V_0(1-\cos\theta)$. If 
$-2V_0\leq{\cal E}<0$, then the pendulum oscillates near one of the minima of the potential, which corresponds to a periodic modulation of the relative phase in real space. The case ${\cal E}=0$ corresponds to the pendulum completing just one
full rotation from $0$ to $2\pi$, or to a single phase soliton connecting $\theta=0$ at $x\to-\infty$ and $\theta=2\pi$ at $x\to\infty$. If ${\cal E}>0$ then the pendulum has enough energy to complete an infinite number of full rotations, 
which corresponds to a soliton lattice. 

It follows from Eq. (\ref{int-of-motion}) that
\begin{equation}
\label{x vs theta}
  x=\int_0^\theta\frac{d\theta'}{\sqrt{2{\cal E}+2V_0(1-\cos\theta')}},
\end{equation}
which implicitly determines $\theta$ as a function of $x$. From the last expression we can immediately obtain the soliton lattice period as a function of $\cal{E}$:
\begin{equation}
\label{SL-period-gen}
  \ell=\int_0^{2\pi}\frac{d\theta}{\sqrt{2{\cal E}+2V_0(1-\cos\theta)}}.
\end{equation}
Focusing on the vicinity of the transition at ${\cal E}=0$ we have
\begin{equation}
\label{SL-period}
  \ell\simeq\frac{2}{\sqrt{V_0}}\ln\left(\frac{V_0}{\cal E}\right),
\end{equation}
with logarithmic accuracy at ${\cal E}\to+0$. The next step is to relate ${\cal E}$ and $\ell$ to the magnetic field $h$.

Using Eq. (\ref{delta f}) and the fact that the phase winding per soliton is equal to $2\pi$, we obtain the following expression for the free energy density:
\begin{equation}
\label{delta f over L}
  \lim_{L_x\to\infty}\frac{1}{L_x}\int_0^{L_x} dx\;\delta f=\frac{\epsilon(\ell)-2\pi h}{\ell},
\end{equation}
where
$$
  \epsilon(\ell)=\int_0^l dx\left[\frac{1}{2}(\nabla_x\theta)^2+V_0(1-\cos\theta)\right]
$$
is the free energy per one cell of the soliton lattice. 

Since the lattice period diverges at the transition, we obtain from Eq. (\ref{delta f over L}) the following expression for the critical field: $h_s=\epsilon(\infty)/2\pi$. 
Introducing the single-soliton energy $\epsilon_1=\epsilon(\infty)$, we recover Eq. (\ref{h_s}). In order to calculate $\epsilon(\ell)$ and $\epsilon_1$, we use the integral of motion, Eq. (\ref{int-of-motion}), and obtain:
\begin{eqnarray}
\label{epsilon_1}
  \epsilon(\ell)&=&\int_0^\ell dx\left[{\cal E}+2V_0(1-\cos\theta)\right]\nonumber\\
	&=&\int_0^{2\pi}d\theta\sqrt{2{\cal E}+2V_0(1-\cos\theta)}-{\cal E}\ell.
\end{eqnarray}
The second line follows after changing the variable, $x\to\theta$, in the first one and using the expression for $dx/d\theta$ from Eq. (\ref{x vs theta}). Putting ${\cal E}=0$, we have $\epsilon_1=8\sqrt{V_0}$.

The lattice period is determined by minimizing the free energy, Eq. (\ref{delta f over L}), with respect to $\ell$. In this way, we obtain:
\begin{equation}
\label{epsilon-derivative}
  \frac{\epsilon-2\pi h}{\ell}=\frac{\partial\epsilon}{\partial\ell}.
\end{equation}
It follows from Eqs. (\ref{epsilon_1}) and (\ref{SL-period-gen}) that $\partial\epsilon/\partial\ell=-{\cal E}$. Substituting this and Eq. (\ref{epsilon_1}) into Eq. (\ref{epsilon-derivative}), we have
\begin{equation}
\label{E vs h}
  \frac{1}{2\pi}\int_0^{2\pi}d\theta\sqrt{2{\cal E}+2V_0(1-\cos\theta)}=h.
\end{equation}
This last equation implicitly determines ${\cal E}$ and, therefore, the lattice period, as functions of the field. The critical field of the soliton transition is given by
\begin{equation}
\label{h_s-V_0}
  h_s=h({\cal E}=0)=\frac{4}{\pi}\sqrt{V_0}.
\end{equation}

Next, using Eqs. (\ref{E vs h}) and (\ref{SL-period}), we find
$$
  \frac{\partial}{\partial{\cal E}}(h-h_s)=\frac{\ell({\cal E})}{2\pi}\simeq\frac{1}{\pi\sqrt{V_0}}\ln\left(\frac{V_0}{\cal E}\right),
$$
near the critical field. Integrating the last equation, we obtain: 
$$
\label{E near h_s}
  {\cal E}(h)\simeq\frac{\pi\sqrt{V_0}(h-h_s)}{\ln[\sqrt{V_0}/\pi(h-h_s)]}
$$
and, therefore,
\begin{equation}
\label{l near h_s}
  \ell(h)\simeq\frac{2}{\sqrt{V_0}}\ln\frac{\sqrt{V_0}}{\pi(h-h_s)},
\end{equation}
with logarithmic accuracy.


\begin{thebibliography}{99}
\bibitem{comment}
The Zeeman interaction is not diagonal in the helicity basis, leading to the interband transitions caused by the magnetic field. The effect of these transitions on the helicity band quasiparticles is 
quadratic in $H$ and is neglected if the Zeeman energy is much smaller than the SO band splitting.
\end{thebibliography}
\end{document}